\documentclass[useAMS,usenatbib]{mnras}
\usepackage{graphicx}
\usepackage{graphics}
\usepackage{amssymb}
\usepackage{amsmath}
\usepackage[T1]{fontenc}
\usepackage[utf8]{inputenc}
\usepackage{pdflscape} 
\usepackage{color} 
\def\beq{\begin{eqnarray}}
\def\eeq{\end{eqnarray}}
\def\bdm{\begin{displaymath}}
\def\edm{\end{displaymath}}
\def\be {\begin{equation}}
\def\ee {\end{equation}}

\newcommand{\td}[1]{\, \mathrm{d} #1 \,}
\newcommand{\intl}{\int\limits}

\newcommand{\muobs}{\mu_{rad}}
\newcommand{\muxi}{\mu_{\xi}}
\newcommand{\rxi}{r_{\xi}}
\newcommand{\cospsi}{\cos{\Psi}}
\newcommand{\TeVCat}{\mbox{TeVCat}}
\title[Attenuation of $\gamma$-rays by starlight]{Attenuation of TeV $\gamma$-rays by the starlight photon field of the host galaxy}
\author[Zacharias, Chen \& Wagner]{Michael Zacharias$^{1,4}$, Xuhui Chen$^{2,3}$, Stefan J. Wagner$^1$ \\
$^1$ Landessternwarte, Universit\"at Heidelberg, K\"onigstuhl, D-69117 Heidelberg, Germany, mzacharias.phys@gmail.com \\
$^2$ Institute of Physics and Astronomy, University of Potsdam, 14476 Potsdam-Golm, Germany \\
$^3$ DESY, Platanenallee 6, 15738 Zeuthen, Germany \\
$^4$ Now at: Centre for Space Research, North-West University, Potchefstroom 2520, South Africa
}
\date{Received 2016 November 04; accepted 2016 November 20; in original form: 2016 August 05 } %
\begin{document}
\maketitle
\begin{abstract}
The absorption of TeV $\gamma$-ray photons produced in relativistic jets by surrounding soft photon fields is a long-standing problem of jet physics. In some cases the most likely emission site close to the central black hole is ruled out because of the high opacity caused by strong optical and infrared photon sources, such as the broad line region. 
Mostly neglected for jet modeling is the absorption of $\gamma$-rays in the starlight photon field of the host galaxy. Analyzing the absorption for arbitrary locations and observation angles of the $\gamma$-ray emission site within the host galaxy we find that the distance to the galaxy center, the observation angle, and the distribution of starlight in the galaxy are crucial for the amount of absorption. We derive the absorption value for a sample of $20$ TeV detected blazars with a redshift $z_r<0.2$. The absorption value of the $\gamma$-ray emission located in the galaxy center may be as high as $20\%$ with an average value of $6\%$.
This is important in order to determine the intrinsic blazar parameters. We see no significant trends in our sample between the degree of absorption and host properties, such as starlight emissivity, galactic size, half-light radius, and redshift. 
While the uncertainty of the spectral properties of the extragalactic background light exceeds the effect of absorption by stellar light from the host galaxy in distant objects, the latter is a dominant effect in nearby sources. It may also be revealed in a differential comparison of sources with similar redshifts. 
\end{abstract}
%
%
\section{Introduction}
Active galactic nuclei (AGN), and especially the blazar sub-class, are well known emitters of TeV $\gamma$-ray emission, which is generally assumed to be produced in the jet by relativistic electrons through inverse Compton scatterings. The production site of the TeV emission is often considered to be within the first few parsecs of the jet \citep[e.g.,][]{gtg09,gea11,bea13}. At these small distances, the influence of the optical broad line region and the infrared dusty torus on the emission of the blazar can be significant for both serving as seed photon fields for the inverse Compton process and at the same time absorbing the $\gamma$-rays produced in the jet. The resulting absorption features have been used to constrain the location of the $\gamma$-ray emission region \citep[e.g.,][]{sp11,dmt12}. This deduction assumes that further absorption within the photon fields of the host galaxy can be neglected.

Beyond the central region the most important photon field is the starlight of the host galaxy, which could further absorb $\gamma$-ray photons of energies between a few hundred GeV and a few TeV.
\citet[hereafter S06]{sea06} derived the optical depth for TeV photons emerging from the central region in the nearby active galaxy Centaurus~A (Cen~A), and obtained a maximum value on the order of $\sim 1.5\%$. As these authors note, this seemingly low value already produces distinctive radiative features. The absorbed radiation is reprocessed in electron-positron cascades and could become visible as a $\gamma$-ray halo of the size of the galaxy. 

Since the absorption by starlight is inevitable irrespective of the type of active galaxy (i.e., whether or not it contains strong central photon sources such as a broad-line region), it is important to understand the potential impact of the host galaxy on the observed spectrum of the active nucleus. Depending on the strength of the absorption this can influence the modeling and interpretation of a large range of quantities, from jet internal parameters, such as the electron spectral index, to parameters of the central photon fields. 

It is important to note that AGN like Cen~A, M~87 and others have been detected at TeV energies despite being no blazars. They potentially exhibit a curved jet, and therefore different amounts of beaming at different positions in the jet. As a consequence the $\gamma$-rays might be emitted non-radially with respect to the galactic center and thus take a longer path through the starlight photon field. It is worth investigating whether the viewing angle has consequences on the amount of absorption.

Furthermore, recent investigations on the low frequency peaked BL Lac object AP Librae have resulted in contrary locations of the $\gamma$-ray emission region. AP Librae has been detected at TeV energies by the H.E.S.S. experiment \citep{hea15} despite its classification, where no TeV emission is expected. The standard blazar one-zone model therefore fails. \cite{hbs15} argue for an emission region close to the central black hole, while \cite{zw16} invoke the observed kpc-scale radio and X-ray jet as the origin of the TeV emission. Given that AP Librae's host galaxy is relatively bright, the different levels of absorption by the starlight photon field expected for these scenarios could help to solve this mystery.

So far, the absorption of TeV radiation by galactic starlight has been treated either for special cases (as in S06) or in an approximated way \citep[as in][]{pc13}. Here we use the approach of S06 and expand it for arbitrary locations and observing angles of the $\gamma$-ray emission site within the host galaxy. In section 2 we derive the respective equations, and perform a parameter study. In section 3 we calculate the degree of absorption for a sample of TeV detected blazars. We conclude in section 4.

The dust between the stars might also emit enough infrared photons to cause absorption at multi-TeV energies. However, we will neglect the effect of the dust, since most AGN emit at less than a few TeV, where the absorption by infrared photons is irrelevant.

%
%
\section{Optical depth due to starlight photons}
\subsection{Derivation of the absorption equation} \label{sec:derequ}
Generally, the optical depth $\tau_{\gamma\gamma}$ is the path integral with respect to the soft photon density $n_{\epsilon}$ folded by the pair production cross section $\sigma_{\gamma\gamma}$. Retracing and expanding the steps of S06 will lead us to the degree of absorption of $\gamma$-rays produced at an arbitrary location within the galaxy observed under an arbitrary observation angle with respect to the radial direction out of the galaxy. We assume that the galaxy is spherically symmetric. This is a valid description for most elliptical galaxies, which are the hosts of most TeV emitting AGN.

The pair production cross-section \citep{bw34,gs67} is given by
\begin{align}
 \sigma_{\gamma\gamma}(\epsilon_{\gamma},\epsilon,\mu) = \frac{3\sigma_T}{16} (1-\beta^2) \nonumber \\
 \times \left[ (3-\beta^4) \ln{\left( \frac{1+\beta}{1-\beta} \right)} - 2\beta (2-\beta^2) \right] \label{eq:sigmagammagamma},
\end{align}
with the Thomson cross-section $\sigma_T = 6.65\times 10^{-25}\,$cm$^2$, the angle-cosine $\mu$ between the direction of the $\gamma$-ray photon with normalized energy $\epsilon_{\gamma} = h\nu_{\gamma} / (m_e c^ 2)$ and the incoming starlight photon with normalized energy $\epsilon = h\nu/ (m_e c^2)$, and the speed of the created electron-positron pair normalized to the speed of light
\begin{align}
 \beta = \left[ 1-\frac{2}{\epsilon_{\gamma}\epsilon (1-\mu)} \right]^{1/2}.
\end{align}
Obviously, for $\epsilon<\epsilon_{thr} = 2/(\epsilon_{\gamma}(1-\mu))$ no electron-positron pair can be created.

The soft starlight photon field acting as an absorber for the $\gamma$ radiation, is assumed to fill the entire galaxy. Its emissivity at any given radius $r$ (as measured from the galactic center) can be approximated as
\begin{align}
 \epsilon j_{\epsilon}(\rho) = j\, g(\epsilon) h(\rho) \label{eq:slemis},
\end{align}
where $j$ marks the monochromatic emissivity of the galaxy, and $\rho = r/r_b$ is the radius normalized to some characteristic radius $r_b$. For the general considerations of this section it is not necessary to specify neither the radial dependence function $h(\rho)$ of the starlight surface brightness nor its spectral distribution $g(\epsilon)$. In principle, these functions can be measured for any galaxy. For the parameter study in section \ref{sec:param} we will use empirical functions.

The differential photon number density $n_{\epsilon}(\rho,\Omega)$ of starlight photons can be calculated from the spectral intensity within a solid angle $\Omega$ by $I_{\epsilon}(\Omega) = c m_ec^2 \epsilon n_{\epsilon}(\rho,\Omega)$. We neglect absorption of starlight photons through intervening dust clouds. This is a valid approximation for elliptical galaxies, which are mostly devoid of dust. Thus, the spectral intensity is the integral of the spectral emissivity along a path $l$ in direction $\Omega$ through the galaxy, $I_{\epsilon}(\Omega) = \int j_{\epsilon}(\rho) \td{l}$. Hence,
\begin{align}
 n_{\epsilon}(\rho,\Omega) = \frac{r_b}{\epsilon^2 m_ec^3} \int \left[ \epsilon j_{\epsilon}(\rho) \right] \td{\eta} \label{eq:nstar},
\end{align}
where the path $l$ of the galactic photons has been normalized to $\eta = l/r_b$.

Consider a $\gamma$-ray emission site at normalized distance $\zeta=z/r_b$ from the galactic center. A $\gamma$-ray photon with energy $\epsilon_{\gamma}$ in the galactic frame is emitted at an angle-cosine $\muobs$ between the radial direction out of the galaxy and the line-of-sight to the observer.\footnote{The angle $\theta_{rad}$ can be identified with the typical observation angle $\theta_{obs}$ if the jet is straight. Otherwise the difference between $\theta_{rad}$ and $\theta_{obs}$ is given by the deviation of the jet from the radial direction.} The photon's path along the line-of-sight out of the galaxy may be parameterized by $\xi = l_{\gamma}/r_b$. The $\gamma$-ray photon escapes the galaxy once it has crossed the normalized terminal radius of the galaxy $R_t = R_{gal}/r_b$, where $R_{gal}$ marks the optical edge of the galaxy. The maximum path length $\xi_t$ of the $\gamma$-ray photon from the production site $\zeta$ to the terminal radius $R_t$ is given by
\begin{align}
 \xi_t = \sqrt{R_t^2 - \zeta^2 (1-\muobs^2)} - \zeta \muobs \label{eq:xit}.
\end{align}
At any point $\xi$ along the path, the $\gamma$-ray photon is at a normalized distance from the galactic center
\begin{align}
 \rxi = \sqrt{\zeta^2 + \xi^2 + 2\zeta\xi\muobs} \label{eq:rxi}.
\end{align}
The angle-cosine between the direction of the $\gamma$-ray photon and the direction to the galactic center is given by
\begin{align}
 \muxi = \frac{\xi + \zeta \muobs}{\rxi} \label{eq:muxi}.
\end{align}

The direction of the incoming starlight photons with respect to the direction of the $\gamma$-ray photon is characterized by the solid angle increment $\td{\Omega} = \td{\mu} \td{\phi}$. While the pair production cross section is independent of the azimuth $\phi$, the starlight photon number density, Eq. (\ref{eq:nstar}), is not. The integral along a path $\eta$ from the interaction site $\xi$ to the outer edge $R_t$ of the galaxy in direction $\Omega$ is bounded by the maximum value
\begin{align}
 \eta_{max} = \sqrt{R_t^2 - \rxi^2 (1-(\cospsi)^2)} + \rxi \cospsi \label{eq:etamax},
\end{align}
with
\begin{align}
 \cospsi = \mu\muxi + \sqrt{1-\mu^2} \sqrt{1-\muxi^2} \cos{\phi} \label{eq:cospsi}.
\end{align}
The path of the starlight photons $\eta$ can be parameterized by the radius $\rho$ from the galactic center as
\begin{align}
 \rho(\eta) = \sqrt{\rxi^2 + \eta^2 - 2\rxi\eta\cospsi} \label{eq:rho}.
\end{align}

With these definitions the optical depth for $\gamma$-ray photons within the starlight photon field of a galaxy originating from a distance $\zeta$ above the galactic center and moving away from the radial direction in direction $\muobs$ becomes
\begin{align}
 \tau_{\gamma\gamma}(\epsilon_{\gamma},\zeta,\muobs) = \intl_0^{\xi_t}\td{\xi} \intl_{\epsilon_{thr}}^{\infty} \td{\epsilon} \oint \td{\Omega} (1-\mu) \nonumber \\
 \times n_{\epsilon} (\zeta,\Omega) \sigma_{\gamma\gamma}(\epsilon_{\gamma},\epsilon,\mu) \nonumber \\
 = \frac{j r_b^2}{m_ec^3} \intl_0^{2\pi} \td{\phi} \intl_{-1}^{1} \td{\mu} (1-\mu) \intl_{\epsilon_{thr}}^{\infty} \td{\epsilon} \sigma_{\gamma\gamma}(\epsilon_{\gamma},\epsilon,\mu) \frac{g(\epsilon)}{\epsilon^2} \nonumber \\
 \times \intl_0^{\xi_t}\td{\xi} \intl_{0}^{\eta_{max}}\td{\eta} h(\rho(\eta)) \label{eq:taugammagammaGEN} .
\end{align}

As can be easily verified, for $\zeta=0$ the result of S06 is recovered. In this case the value of $\muobs$ is irrelevant. As a matter of fact, this equation is generally true for any absorber, since no assumption on the spectral and spatial distribution of the soft photon field has been made, as long as the $\gamma$-ray photon is within the soft photon field. If the $\gamma$-ray photon is outside the soft photon field, a factor $\mu$ needs to be multiplied to the integrand, and the limits of the $\mu$-integral must be adjusted.

%
\subsection{Parameter study} \label{sec:param}
%
In order to quantify the preceding result, specifications need to be made with respect to the starlight radial dependence function $h(\rho)$ and the starlight spectral distribution $g(\epsilon)$.

The radial dependence function $h$ can be derived from the galactic surface brightness $S$ as $h(\rho)\propto \rho^{-1}S(\rho)$ normalized to $h(1)=1$. We describe the surface brightness by the de Vaucouleurs' profile, yielding the radial dependence function
\begin{align}
 h(\rho) = \rho^{-1} \exp{\left\{ -7.67 \left( \rho^{1/4}-1 \right) \right\}} \label{eq:deVaucprofile}
\end{align}
The radius $\rho$ is normalized to the half-light radius $r_h$. This profile describes well elliptical galaxies, which are the majority of blazar hosts, and measured values are provided for a large number of galaxies. The main weakness of this profile is its inability to describe the nuclear region of the galaxy, where profiles tend to flatten.

The spectral distribution of the starlight of powerful elliptical galaxies is well approximated by the template model of \cite{sea98}. Using only the direct stellar emission, restricts the energy range of the template spectrum between $\epsilon_{min}=10^{-7}$ and $\epsilon_{max}=10^{-5}$. Then, the template model can be approximated as
\begin{align}
 g(\epsilon) = 
 \begin{cases}
   g_a\epsilon^{2.44} & \mbox{for} \, 10^{-7.0}\leq \epsilon< 10^{-5.8} \\ 
   g_b\epsilon^{-0.57} & \mbox{for} \, 10^{-5.8}\leq \epsilon< 10^{-5.3} \\ 
   g_c\epsilon^{-5.74} & \mbox{for} \, 10^{-5.3}\leq \epsilon\leq 10^{-5.0} 
 \end{cases}
  \label{eq:spectemp} .
\end{align}
The constants $g_i$ must be adjusted to the frequency of the monochromatic emissivity $j$ that is chosen in the analysis.

With these specifications, we compute the optical depth in a galactic photon field given by Eq. (\ref{eq:taugammagammaGEN}) using a Monte Carlo scheme. 
As a benchmark case, we construct an average galaxy out of the sample presented in section \ref{sec:appl}. The parameters are $r_h = 2.36\times 10^{22}\,$cm, $R_{gal} = 25.81\,$kpc, $j_R = 7.73\times 10^{-26}\,$erg/cm$^3$/s, $g_a=2.34\times 10^{14}$, $g_b = 8.11\times 10^{-4}$, and $g_c = 3.11\times 10^{-31}$. This results in a total luminosity of $L_R = 4.71\times 10^{44}\,$erg/s.
\begin{figure}
\centering
\includegraphics[width=0.48\textwidth]{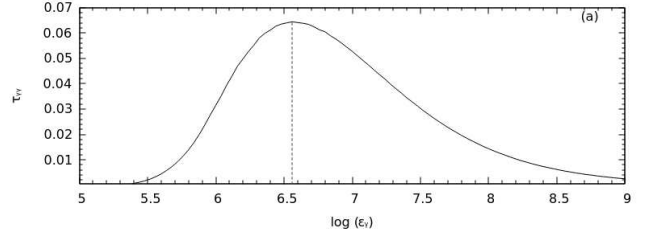}
\caption{Optical depth $\tau_{\gamma\gamma}$ as a function of the normalized $\gamma$-ray energy $\epsilon_{\gamma}$. Here, we use the parameters: galactocentric distance $\zeta=0$, half-light radius $r_h=2.36\times 10^{22}\,$cm, observation angle-cosine $\muobs=1$, and normalized galactic radius $R_t = 3.39$. The energy $\epsilon_{\gamma} = 3.63\times 10^{6}$, marked by the gray dotted line, indicates the benchmark value in Fig. \ref{fig:tau}.}
\label{fig:eps}
\end{figure}
\begin{figure*}
\begin{minipage}{0.49\linewidth}
\centering \resizebox{\hsize}{!}
{\includegraphics{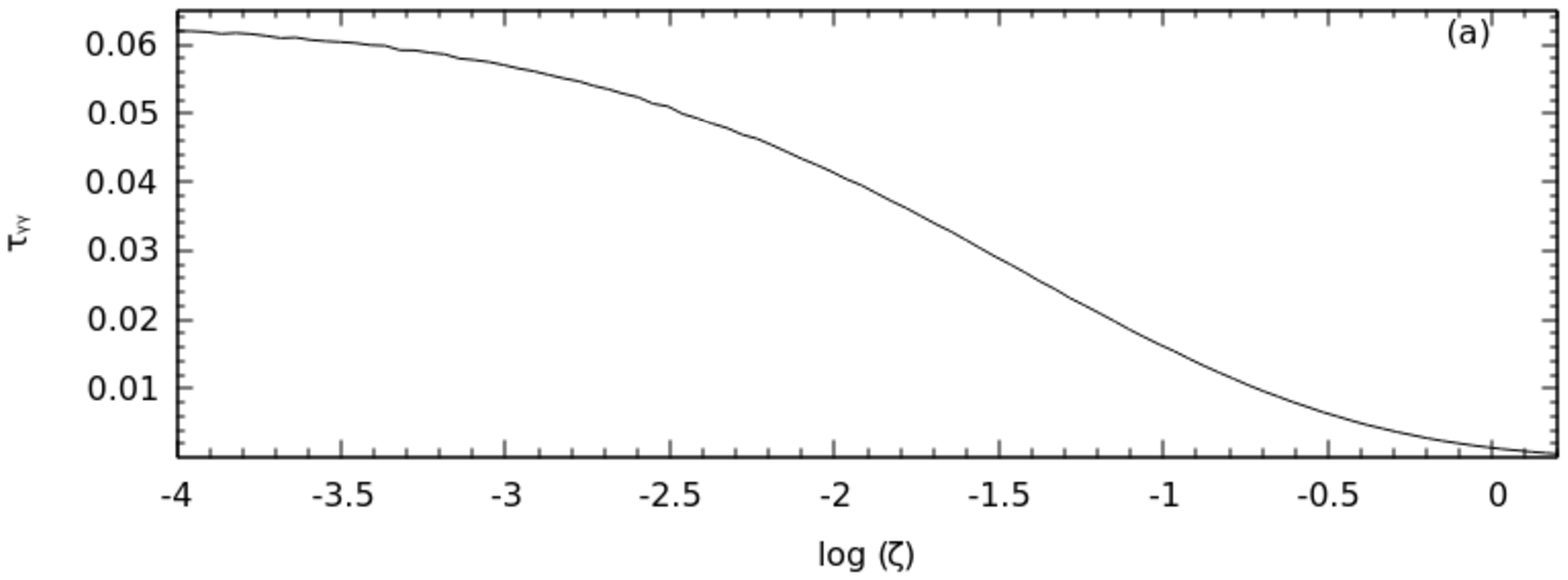}}
\end{minipage}
\hspace{\fill}
\begin{minipage}{0.49\linewidth}
\centering \resizebox{\hsize}{!}
{\includegraphics{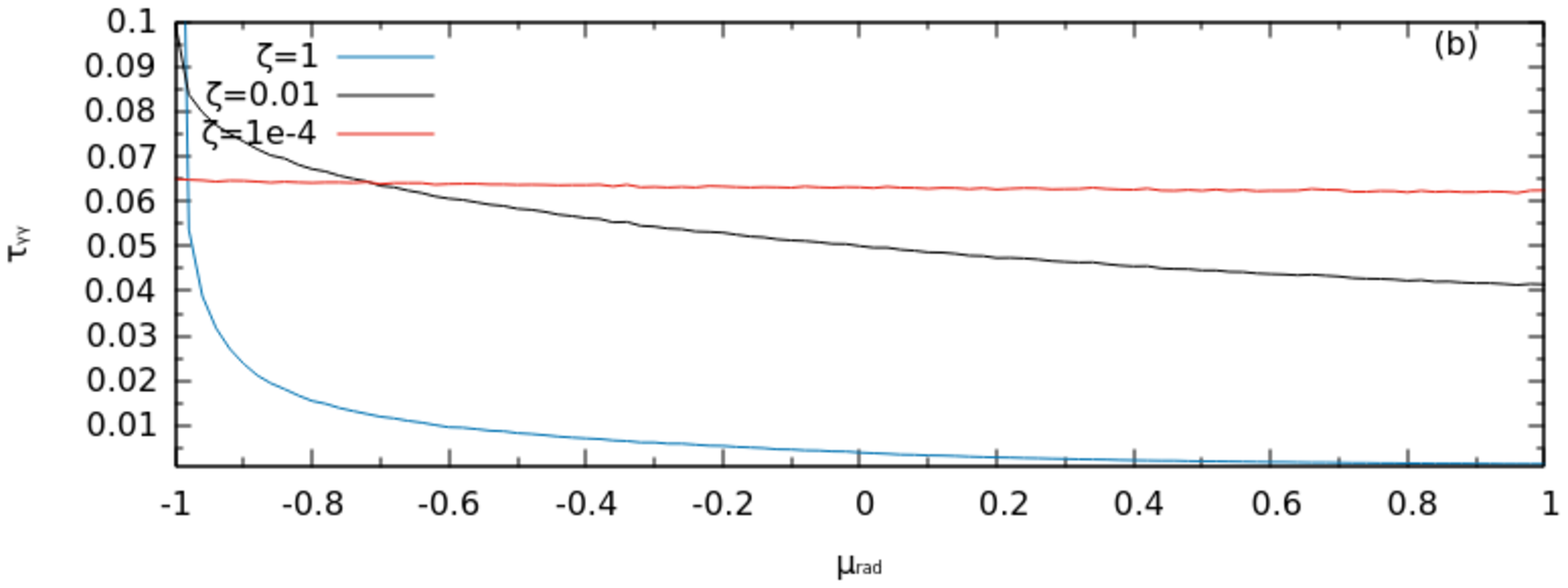}}
\end{minipage}
\newline
\begin{minipage}{0.49\linewidth}
\centering \resizebox{\hsize}{!}
{\includegraphics{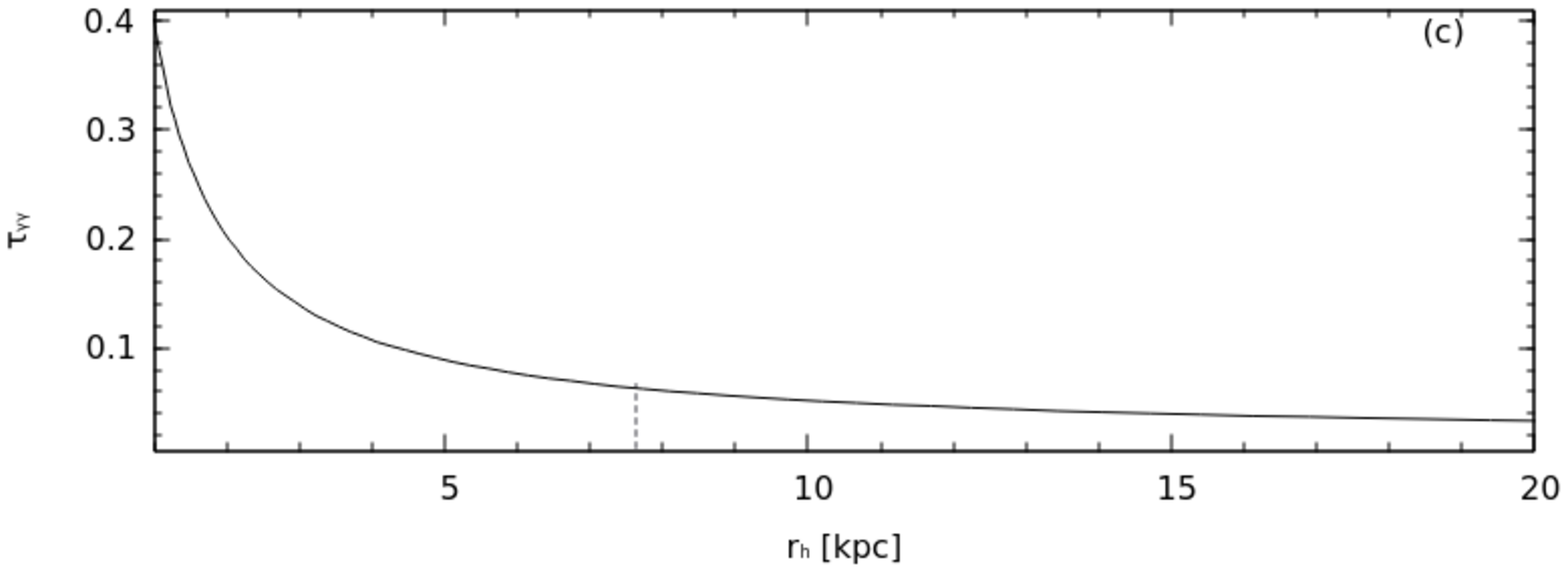}}
\end{minipage}
\hspace{\fill}
\begin{minipage}{0.49\linewidth}
\centering \resizebox{\hsize}{!}
{\includegraphics{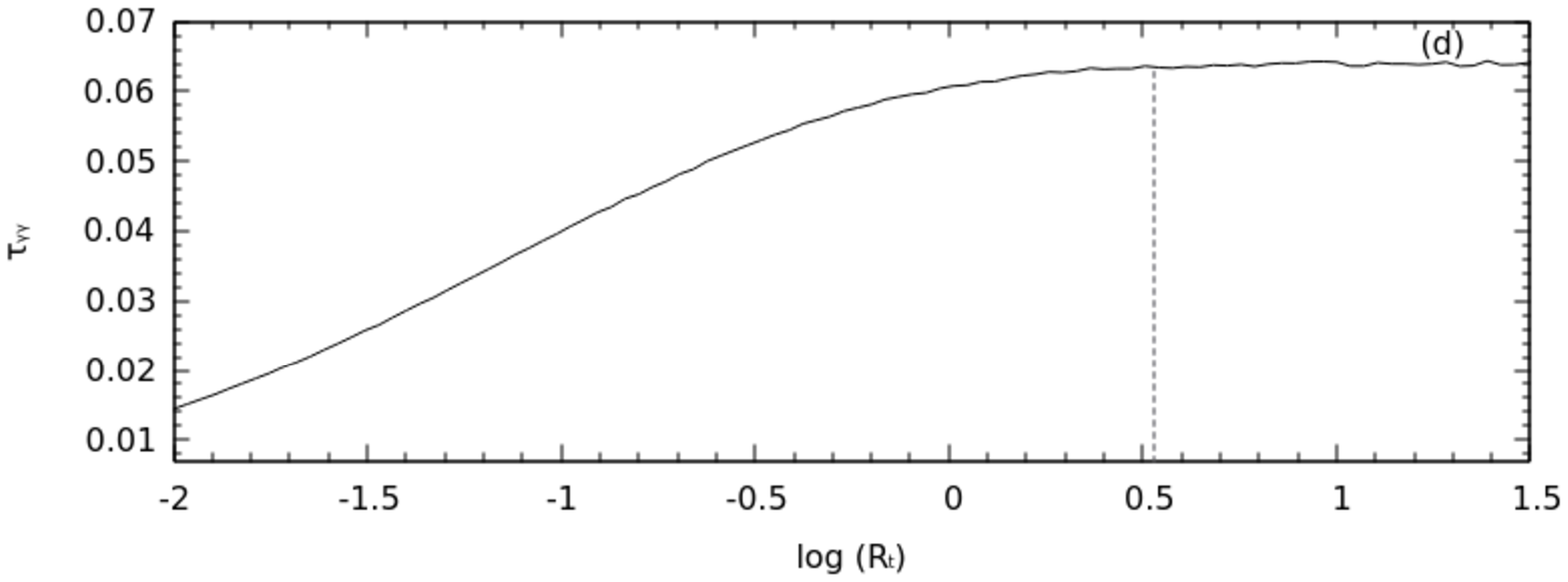}}
\end{minipage}
\caption{Optical depth $\tau_{\gamma\gamma}$ as a function of different parameters at fixed energy $\epsilon_{\gamma}=3.63\times 10^{6}$. 
(a) Opacity $\tau_{\gamma\gamma}$ as a function of galactocentric distance $\zeta$ with the parameters: half-light radius $r_h=2.36\times 10^{22}\,$cm, observation angle-cosine $\muobs=1$, and normalized galactic radius $R_t = 3.39$. 
(b) Opacity $\tau_{\gamma\gamma}$ as a function of observation angle-cosice $\muobs$ for three values of galactocentric distance $\zeta$ as indicated with the parameters: half-light radius $r_h=2.36\times 10^{22}\,$cm, and normalized galactic radius $R_t = 3.39$. 
(c) Opacity $\tau_{\gamma\gamma}$ as a function of half-light radius $r_h$ with the parameters: galactocentric distance $\zeta=0$, observation angle-cosine $\muobs = 1$, and galactic radius $R_{gal} = 8.0\times 10^{22}\,$cm. 
(d) Opacity $\tau_{\gamma\gamma}$ as a function of normalized galactic radius $R_t$ with the parameters: galactocentric distance $\zeta=0$, half-light radius $r_h=2.36\times 10^{22}\,$cm, and observation angle-cosine $\muobs=1$.
The gray dotted line (c,d) gives the benchmark value.}
\label{fig:tau}
\end{figure*} 

In Fig. \ref{fig:eps} we show the dependence of the optical depth $\tau_{\gamma\gamma}$ as a function of $\gamma$-ray energy $\epsilon_{\gamma}$. The plot parameters are set to $\zeta=0$, $\muobs=1$, and $R_t = 3.39$. These values influence the magnitude of absorption, but not the spectral form. The distribution of the optical depth around the maximum is broad, so the absorption is effective for roughly an order of magnitude in energy. The maximum is attained at $\epsilon_{\gamma} \sim 3.63\times 10^6$ corresponding to an energy of $\sim 2\,$TeV. 

In order to discuss the dependencies of $\tau_{\gamma\gamma}$ on the other parameters, we fix the energy at the maximum value of $\epsilon_{\gamma}=3.63\times 10^{6}$. The dependencies on the distance $\zeta$ to the core, the observation angle $\muobs$, the half-light radius $r_h$, and the galactic size $R_t$ are shown in Fig. \ref{fig:tau}.

To visualize the dependence on the core distance $\zeta$ (Fig. \ref{fig:tau} (a)), we set $\muobs=1$ and $R_t=3.36$. As expected, $\tau_{\gamma\gamma}(\zeta)$ is a monotonically decreasing function. Nevertheless, for an emission region at $\zeta=0.025$ the optical depths is still at $50\%$ of the central value. In our template galaxy, this corresponds to a distance of $\sim 190\,$pc from the black hole. For $\zeta>1$, i.e. an emission region located beyond the half-light radius, the absorption becomes irrelevant compared to the inner regions, since the absorption is smaller than $3\%$ of the central value. Thus, most of the absorption happens in the inner regions of the galaxy.

In Fig. \ref{fig:tau} (b) we show the dependence on the observation angle $\muobs$ for three values of the distance $\zeta$, namely $\zeta=1$ (blue line), $\zeta=0.01$ (black line), and $\zeta = 0.0001$ (red line). The galactic size is again set to $R_t=3.36$. In all cases, the lowest absorption takes place for $\muobs=1$, while for $\muobs=-1$ the absorption can be several times larger. This is expected, since for $\muobs=1$ the photons take the shortest route out of the galaxy, while for $\muobs=-1$ they travel the longest possible way and have to cross the bright central region of the galaxy. However, for a large range of observation angles the absorption does not change significantly, since the bright central region is only crossed for very large observation angles ($\muobs\rightarrow -1$). The magnitude of the change for different viewing angles depends sensitively on $\zeta$. For $\zeta\rightarrow 0$ the curve becomes flat, while for $\zeta\rightarrow R_t$ the ratio between the highest and lowest value increases strongly. This also means that a counter-jet in an AGN becomes more difficult to observe in the TeV domain, since apart from the strong de-beaming effect its photons are more strongly attenuated than the photons of the approaching jet. This is especially true for regions with larger distances to the core. This point might become more important once the angular resolution of $\gamma$-ray observatories is good enough to resolve jet structures. Nevertheless, it is obvious that for most AGN observations the viewing angle only has a minor influence on the degree of absorption by the starlight photon field.

For the third case, $\tau_{\gamma\gamma}$ as a function of $r_h$ (Fig. \ref{fig:tau} (c)), $R_{gal}$ is kept fixed at $8.0\times 10^{22}\,$cm, i.e. $R_{t}$ decreases for increasing $r_h$. The emissivity is normalized by the total luminosity of the benchmark case. Hence, for all values of the half-light radius, the integrated flux is the same. Since the relationship between the luminosity and the emissivity scales as $r_h^{-3}$ and the absorption in Eq. (\ref{eq:taugammagammaGEN}) with $r_h^2$, the fixed luminosity results roughly in an inverse dependence of the absorption with the half-light radius. This can be seen in Fig. \ref{fig:tau} (c), where the emission region is located again at the galactic center, $\zeta=0$. The inverse relation of the absorption with the half-light radius is reasonable, given that by definition half of the light is emitted within the half-light radius. Therefore, with increasing half-light radius the starlight within the half-light radius becomes more diluted, dropping the degree of absorption.

The dependence of the absorption on the galactic size $R_t$ is shown in Fig. \ref{fig:tau} (d). This case displays the complementary trend of Fig. \ref{fig:tau} (a). Given that the normalization of the profile is the same for each value of $R_t$ (since $r_h$ is fixed), the resulting curve demonstrates how different integration lengths from the emission region influence the absorption. Naturally, for increasing galactic size the absorption increases, since the $\gamma$-rays have to pass through more light. However, due to the nature of the radial profile, namely that at larger distances from the core the starlight density decreases, at radii $\geq r_h$ the influence of the added starlight becomes negligible. This is shown by the curve flattening out and seemingly approaching an asymptotic value. This seems counter-intuitive, given that the half-light radius by definition contains only half of the total luminosity of the galaxy. However, the starlight density at larger radii is so low that the chance of a $\gamma$-$\gamma$ interaction rapidly decrease. Hence, the outer parts of the galaxy beyond the half-light radius only have a minor influence on the absorption.

Summarizing, a high value of the total luminosity and a low value of the half-light radius favor a high degree of absorption. As long as the galactic size is greater than the half-light radius, the former does not influence the result. Naturally, a larger distance of the emission region from the galactic center reduces the degree of absorption, while a larger viewing angle increases it. However, for most observation angles, the difference is low and can be safely ignored.
%
\subsection{Modeling} \label{sec:mod}
\begin{figure*}
\begin{minipage}{0.49\linewidth}
\centering \resizebox{\hsize}{!}
{\includegraphics{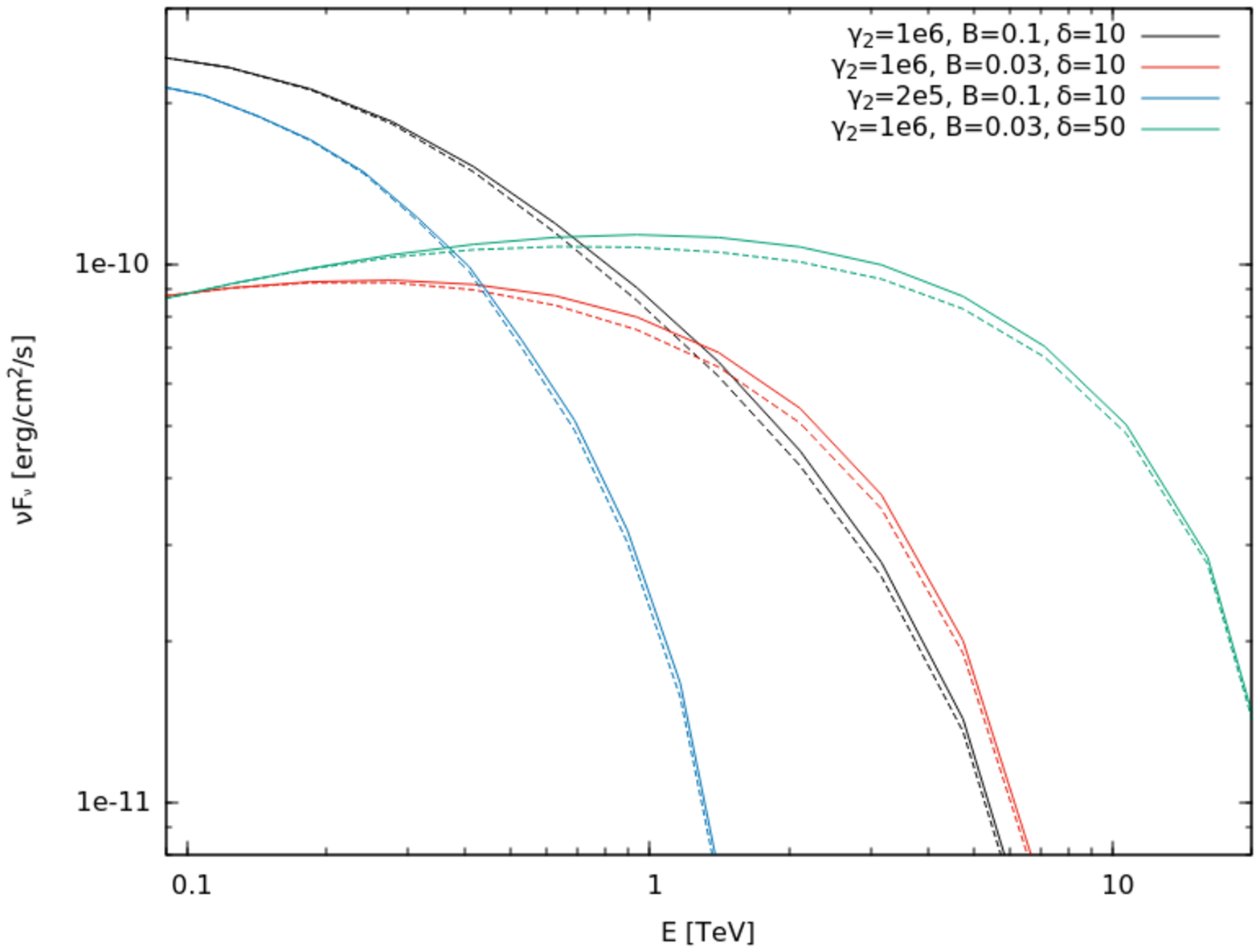}}
\end{minipage}
\hspace{\fill}
\begin{minipage}{0.49\linewidth}
\centering \resizebox{\hsize}{!}
{\includegraphics{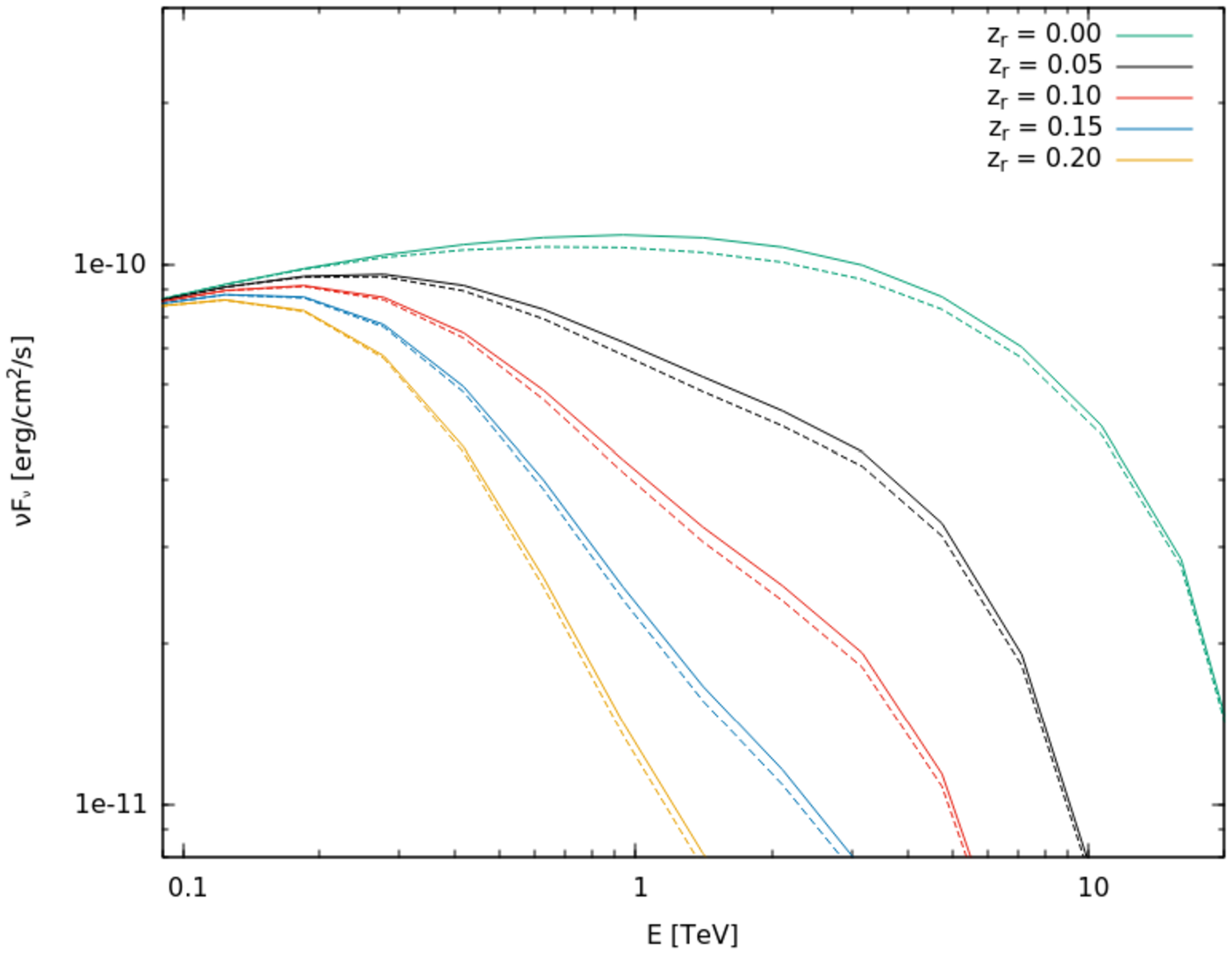}}
\end{minipage}
\caption{\textit{Left:} Intrinsic model SSC spectral energy distributions $\nu F_{\nu}$ as a function of observed photon energy $E$ without (solid lines) and with absorption (dashed lines) due to host galactic starlight. Colors mark different intrinsic parameter sets as given in the legend for maximum electron Lorentz factor $\gamma_2$, magnetic field $B$ in Gauss, and Doppler factor $\delta$. \textit{Right:} Intrinsic model SSC spectral energy distributions $\nu F_{\nu}$ as a function of observed photon energy $E$ for $\gamma_2 = 10^6$, $B = 0.03\,$G, and $\delta = 50$ for different redshifts $z_r$ (solid lines), and with host galactic absorption (dashed lines).}
\label{fig:mod}
\end{figure*} 
The maximum degree of absorption by the starlight of the benchmark galaxy on the blazar emission is on the order of $6\%$. However, whether this effect is detectable does not only depend on the accuracy of the observed spectra, but also on the intrinsic properties of the blazar source and its redshift. We calculated some model synchrotron-self Compton (SSC) spectra with different sets of parameters, to qualitatively study the effect of the absorption.

We have used the code by \cite{bea13}, where the broken power-law electron distribution is derived self-consistently, i.e. the break electron Lorentz factor is derived from the injection parameters, and the spectral index above the break is fixed to $s+1$ with $s$ being the injection spectral index. The electron distribution is bounded by a lower and upper electron Lorentz factor; $\gamma_1$ and $\gamma_2$, respectively. The break electron Lorentz factor depends on the magnetic field $B$ and the size $R$ of the emission region. The blazar emission is beamed into the galactic rest frame, which is described by the Doppler factor $\delta$. For the model spectra in Fig. \ref{fig:mod} left we fixed $s=2.2$, $\gamma_1=10$, and $R = 10^{15}\,$cm. The varied parameters are given in the legend of Fig. \ref{fig:mod} left.

The absorption effect is more pronounced for harder spectra in the TeV regime. For spectra cutting off below $1\,$TeV (blue curve), the absorption cannot be detected. For sources cutting off at a few TeV, the potential for seeing the absorption depends strongly on the steepness of the spectrum. The black and the red curve almost cut off at the same energy, and yet the absorption is easier to detect in the red curve, since its flatter slope at $\sim 2\,$TeV enhances the visibility of the absorption compared to the black curve. In case of the green curve, where the maximum of the SSC flux is attained at a few TeV, the absorption is clearly visible, since the highest degree of absorption roughly coincides with the highest SSC flux. In this case, the quality of the observed spectrum is the biggest uncertainty.

So far we have neglected the influence of the extragalactic background light (EBL), which softens VHE spectra. This redshift-dependent effect is illustrated in Fig. \ref{fig:mod} right. As an intrinsic template we have chosen the spectrum shown in green in the left-hand diagram of the figure, where the host galactic absorption is most obvious. The solid lines in Fig. \ref{fig:mod} right mark the intrinsic spectrum for a source at the indicated redshift including the absorption by the EBL with the model of \cite{fea08}. The dashed lines illustrate the additional absorption of the host galaxy. For increasing redshifts the EBL absorbed spectra become softer, which makes the detection of the galactic absorption effect more difficult. Thus, a high redshift has a similar effect as a low intrinsic cut-off. 

The level of the EBL is not precisely known. The values of $\tau_{EBL}$ are subject to non-negligible uncertainty \citep[e.g.,][]{sea16}. Hence, the effect of EBL contributions cannot be corrected in observed data. At low redshifts however, these results indicate that a detection of the starlight absorption should be dominant especially for so-called extreme HBL, where the maximum of the inverse Compton component is attained at a few TeV. 

%
\section{Application} \label{sec:appl}
\begin{table*}
\caption{Model independent source parameters.}
\label{tab:source}
\begin{tabular}{lccccccccl}
\hline
Name	& Type	& RA (J2000)	& Dec (J2000)	& $z_r$	& $D_L$	& Size scale	& $A_R$	& \multicolumn{2}{c}{$R_{gal}$}	\\
	& 	& 		&		&	& \tiny{[Mpc]}	& \tiny{[kpc/arcsec]}	&	& \tiny{[kpc]}		& \tiny{Ref}	\\
(1)	& (2)	& (3)		& (4)		& (5)	& (6)	& (7)		& (8)	& \multicolumn{2}{c}{(9)} \\
\hline
AP Librae	& LBL	& 15h17m41.8s	& -24d22m19s	& 0.049	& 219	& 0.965	& 0.299	& 42.46	& [3] \\
W Comae		& IBL	& 12h21m31.7s	& $+$28d13m59s	& 0.102	& 473.3	& 1.89	& 0.051	& 14.74	& [2] \\
BL Lac		& IBL	& 22h02m43.3s	& $+$42d16m40s	& 0.069	& 312.9	& 1.328	& 0.713	& 27.89	& [3] \\
1ES 0229$+$200	& HBL	& 02h32m48.6s	& $+$20d17m17s	& 0.14	& 666.2	& 2.458	& 0.295	& 51.62	& [1] \\
RBS 0413	& HBL	& 03h19m51.8s	& $+$18d45m34s	& 0.19	& 932.7	& 3.193	& 0.286	& 31.61	& [2] \\
1ES 0347-121	& HBL	& 03h49m23.2s	& -11d59m27s	& 0.188	& 921.8	& 3.167	& 0.1	& 19.64	& [1] \\
PKS 0548-322	& HBL	& 05h50m40.5s	& -32d16m17s	& 0.069	& 312.9	& 1.328	& 0.077	& 24.76	& [2] \\
RGB J0710$+$591	& HBL	& 07h10m30.1s	& $+$59d08m20s	& 0.125	& 589.1	& 2.256	& 0.083	& 31.57	& [2] \\
1ES 0806$+$524	& HBL	& 08h09m49.2s	& $+$52d18m58s	& 0.138	& 655.9	& 2.455	& 0.097	& 42.96	& [1] \\
Mrk 421		& HBL	& 11h04m27.3s	& $+$38d12m32s	& 0.031	& 136.7	& 0.624	& 0.033	& 13.10	& [2] \\
Mrk 180		& HBL	& 11h36m26.4s	& $+$70d09m27s	& 0.045	& 200.6	& 0.89	& 0.028	& 16.02	& [2] \\
1ES 1215$+$303	& HBL	& 12h17m52.1s	& $+$30d07m01s	& 0.13	& 614.6	& 2.334	& 0.051	& 19.61	& [2] \\
1ES 1218$+$304	& HBL	& 12h21m21.9s	& $+$30d10m37s	& 0.182	& 889.2	& 3.085	& 0.044	& 33.93	& [1] \\
H 1426$+$428	& HBL	& 14h28m32.6s	& $+$42d40m21s	& 0.129	& 609.5	& 2.318	& 0.027	& 16.34	& [2] \\
Mrk 501		& HBL	& 16h53m52.2s	& $+$39d45m37s	& 0.034	& 150.3	& 0.682	& 0.041	& 42.97	& [3] \\
1ES 1727$+$502	& HBL	& 17h28m18.6s	& $+$50d13m10s	& 0.055	& 246.9	& 1.076	& 0.064	& 14.25 & [2] \\
1ES 1959$+$650	& HBL	& 19h59m59.8s	& $+$65d08m55s	& 0.048	& 214.4	& 0.947	& 0.357	& 18.94	& [2] \\
PKS 2005-489	& HBL	& 20h09m25.4s	& -48d49m54s	& 0.071	& 322.5	& 1.363	& 0.122	& 19.08	& [4] \\
1ES 2344$+$514	& HBL	& 23h47m04.8s	& $+$51d42m18s	& 0.044	& 196	& 0.872	& 0.458	& 13.95	& [1] \\
H 2356-309	& HBL	& 23h59m07.9s	& -30d37m41s	& 0.165	& 797.7	& 2.85	& 0.029	& 25.65	& [2] \\
\hline
\end{tabular}
\newline
\begin{flushleft}
Columns:  (1) Source Name; (2) Source Type; (3) Right Ascension; (4) Declination; (5) Redshift; (6) Luminosity distance; (7) Arcsec to kpc conversion factor; (8) R-band Galactic extinction from the NED; (9) Radius of the galaxy and reference. \\
References: [1] \cite{fea99}; [2] NED; [3] \cite{pea02}; [4] \cite{sea00}. \\
\end{flushleft}
\end{table*}
\begin{table}
\caption{Source parameters for the de Vaucouleurs' profile.}
\label{tab:deVauc}
\begin{tabular}{lcclcc}
\hline
Name	& $r_h$ & $\mu_R$ & Ref. & $\log{j_R}$ & $\tau_{\gamma\gamma}$ \\
	& \tiny{[kpc]}	& \tiny{[mag/arcsec$^2$]} & 	& \tiny{[erg/cm$^3$/s]}	& 	\\
(1)	& (2) & (3) & (4) & (5)	& (6)	\\
\hline
AP Librae	& 6.48	& 21.81	& [3]	& -25.15	& 0.044	\\
W Comae		& 3.96	& 21.36	& [2]	& -24.85	& 0.032	\\
BL Lac		& 8.80	& 22.57	& [3]	& -25.41	& 0.043	\\
1ES 0229$+$200	& 9.83	& 21.06	& [1]	& -25.02	& 0.132	\\
RBS 0413	& 10.37	& 22.56	& [4]	& -25.65	& 0.034	\\
1ES 0347-121	& 3.96	& 20.63	& [4]	& -24.54	& 0.066	\\
PKS 0548-322	& 9.36	& 21.89	& [4]	& -25.42	& 0.047	\\
RGB J0710$+$591	& 11.50	& 22.44	& [2]	& -25.73	& 0.035	\\
1ES 0806$+$524	& 4.17	& 19.48	& [1]	& -24.10	& 0.201	\\
Mrk 421		& 6.86	& 21.33	& [2]	& -25.08	& 0.056	\\
Mrk 180		& 6.23	& 21.73	& [2]	& -25.20	& 0.035	\\
1ES 1215$+$303	& 19.48	& 23.31	& [4]	& -26.32	& 0.025	\\
1ES 1218$+$304	& 6.69	& 20.88	& [1]	& -24.88	& 0.085	\\
H 1426$+$428	& 8.57	& 21.83	& [2]	& -25.38	& 0.043	\\
Mrk 501		& 11.72	& 22	& [3]	& -25.58	& 0.053	\\
1ES 1727$+$502	& 3.38	& 21.08	& [4]	& -24.66	& 0.036	\\
1ES 1959$+$650	& 4.82	& 21.59	& [4]	& -24.90	& 0.042	\\
PKS 2005-489	& 7.70	& 21.3	& [4]	& -25.08	& 0.070	\\
1ES 2344$+$514	& 5.05	& 20.24	& [1]	& -24.34	& 0.167	\\
H 2356-309	& 5.27	& 21.08	& [4]	& -24.87	& 0.054	\\
\hline
\end{tabular}
\newline
\begin{flushleft}
Columns:  (1) Name; (2) Half-light radius in kpc; (3) R-band surface brightness at half-light radius; (4) Reference of the previous values; (5) Logarithm of the emissivity; (6) Optical depth for $\zeta = 0$ at $E\sim 2\,$TeV. \\ 
References:  [1] \cite{fea99}; [2] \cite{nea03}; [3] \cite{pea02}; [4] \cite{uea00}. \\
\end{flushleft}
\end{table}
%
\subsection{Source selection} \label{sec:sose}
The template galaxy of the previous section is an average of a sample of hosts of currently detected TeV emitting blazars. In this section we derive the degree of absorption for each galaxy of this sample. We selected blazars listed in the \TeVCat\footnote{\url{http://tevcat.uchicago.edu/}} with a redshift $z_r\leq 0.2$. For galaxies with a larger redshift, the potentially emitted TeV radiation is absorbed due to attenuation by the EBL (see Fig. \ref{fig:mod} right). Only $20$ out of $37$ sources discovered so far are hosted by galaxies which have been characterized sufficiently well (see Tab \ref{tab:source}).

The majority of sources belongs to the class of high frequency peaked BL Lac objects (HBL), while only a small number of sources are intermediate frequency peaked BL Lac objects (IBL), or low frequency peaked BL Lac objects (LBL). This reflects the detection statistics of blazars by the current Cherenkov experiment generation. There are no flat spectrum radio quasars in our sample, since these are located at redshifts $z_r > 0.2$.

The cosmological parameters, i.e. luminosity distance $D_L$ and size scale, were computed with the online tool CosmoCalc\footnote{\url{http://www.astro.ucla.edu/~wright/CosmoCalc.html}} \citep{w06} using a standard flat cosmology with $H_0 = 69.6\,$km/s/Mpc and $\Omega_M = 0.286$.

For all galaxies in our sample the surface brightness profiles are described by a de Vaucouleurs' fit. The profile parameters are the half-light radius $r_h$ and the surface brightness at that radius $\mu_R$. Both values are listed in Tab.~\ref{tab:deVauc}. 

The derivation of the starlight emissivity $j$ is presented in appendix \ref{app:emis} and yields in the R-band
\begin{align}
 j_R = 2.32\times 10^{-5} r_h^{-1} 10^{0.4(25.992-(\mu_R-A_R))} \label{eq:emis}.
\end{align}
Here, $A_R$ is the foreground R-band extinction in the Milky Way. These values are listed in Tab. \ref{tab:source}. 
Using Eq. (\ref{eq:emis}), we derive the emissivities of each source, which are given in Tab. \ref{tab:deVauc}.

%
\subsection{Results} \label{sec:dvp}
\begin{figure*}
\begin{minipage}{0.49\linewidth}
\centering \resizebox{\hsize}{!}
{\includegraphics{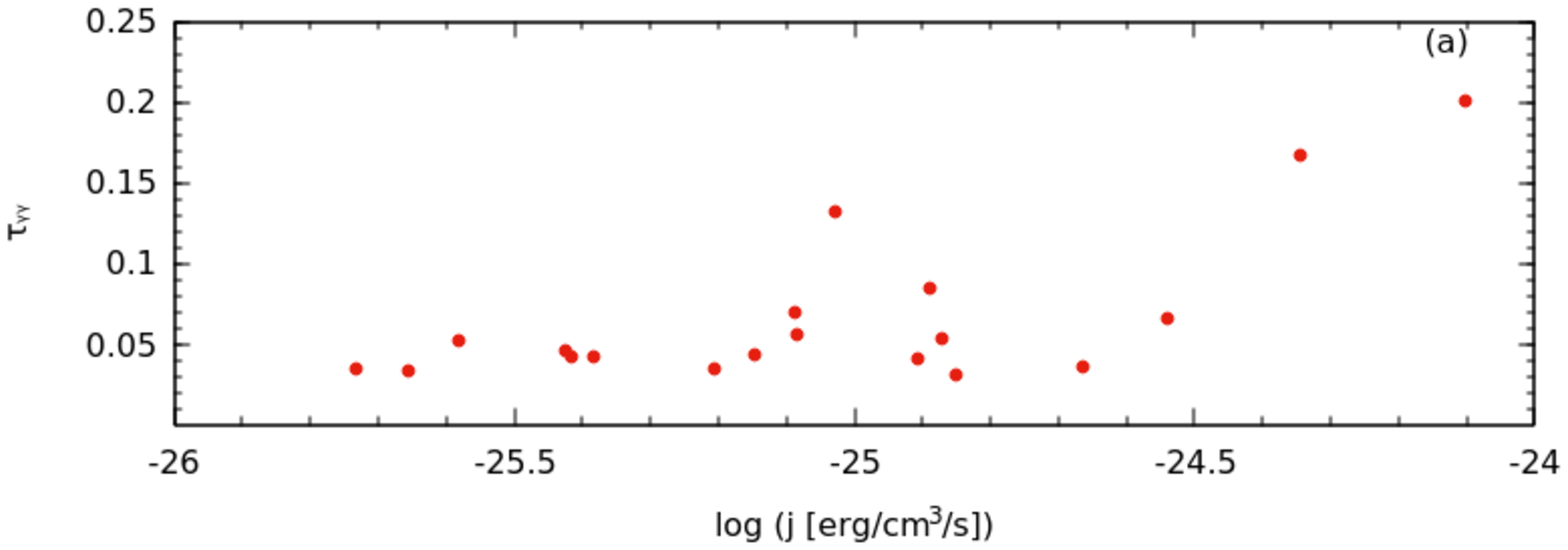}}
\end{minipage}
\hspace{\fill}
\begin{minipage}{0.49\linewidth}
\centering \resizebox{\hsize}{!}
{\includegraphics{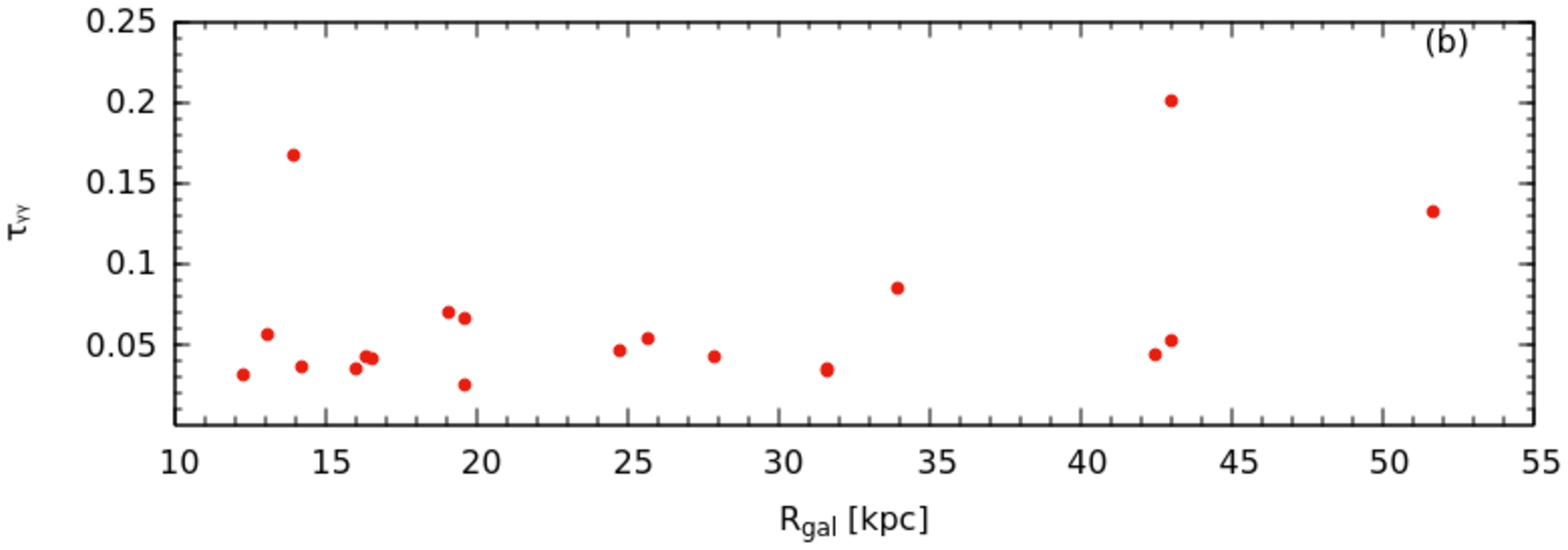}}
\end{minipage}
\newline
\begin{minipage}{0.49\linewidth}
\centering \resizebox{\hsize}{!}
{\includegraphics{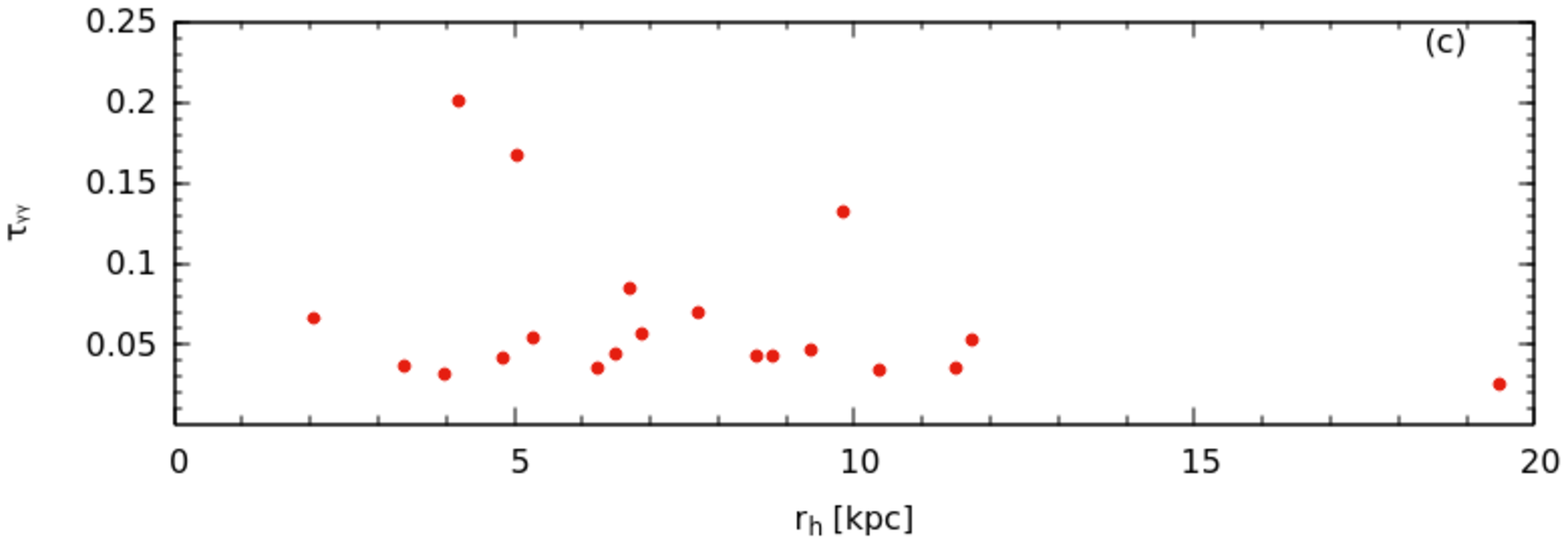}}
\end{minipage}
\hspace{\fill}
\begin{minipage}{0.49\linewidth}
\centering \resizebox{\hsize}{!}
{\includegraphics{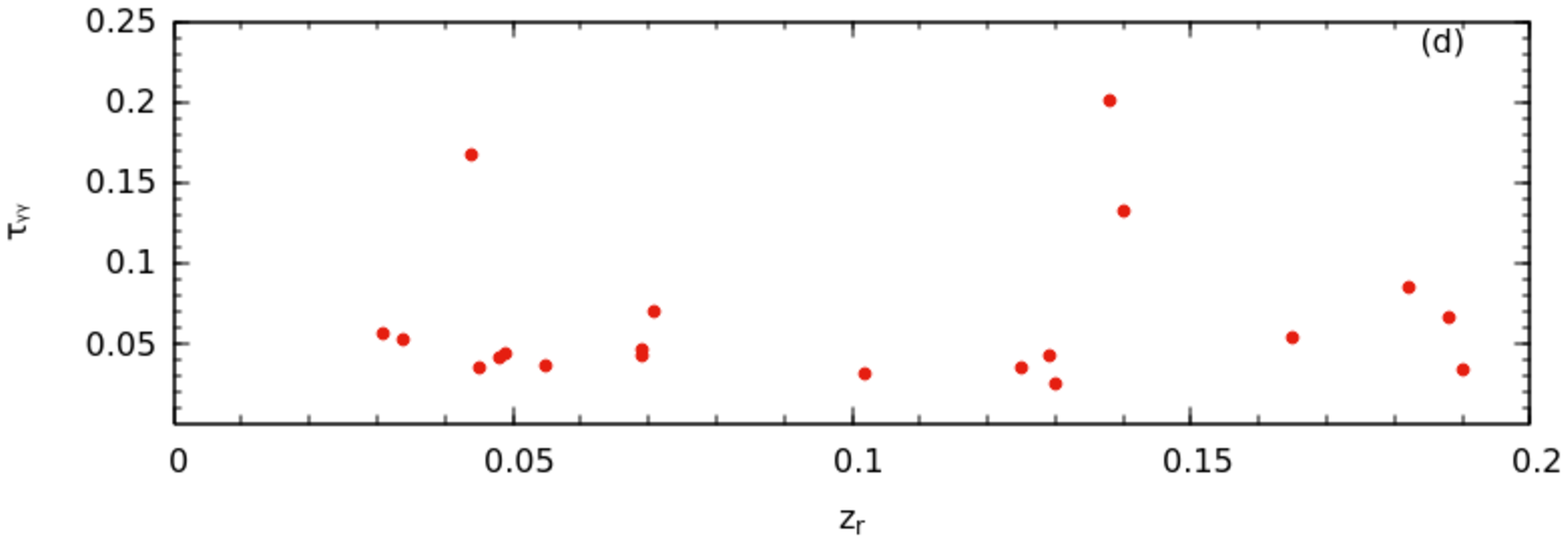}}
\end{minipage}
\caption{Optical depth $\tau_{\gamma\gamma}$ as a function of (a) host emissivity $j$, (b) galactic radius $R_{gal}$, (c) half-light radius $r_{h}$, and (d) redshift $z_r$. The red points mark the optical depth for a $\gamma$-ray emission region at galactocentric distance $\zeta=0$.} 
\label{fig:vaucsource}
\end{figure*} 

Column (6) in Tab. \ref{tab:deVauc} lists the optical depths for an emission region at the galactic center, $\zeta = 0$. The optical depths are on the order of a few percent with an average value $\overline{\tau_{\gamma\gamma}} \sim 0.06$. In order to investigate possible relations of the optical depth with host parameters, we show the respective cases in Fig. \ref{fig:vaucsource}.

In plot (a) of Fig. \ref{fig:vaucsource} the optical depth is displayed versus the emissivity. There seems to be a trend of higher optical depth with higher emissivity. This is reasonable given that a larger emissivity results in a higher density of starlight photons. This trend becomes much weaker, if we remove the three galaxies with the highest absorption values. Hence, while emissivity linearly effects the absorption degree according to Eq. (\ref{eq:taugammagammaGEN}), the interplay with other parameters of the galaxy is complex and results in similar degrees of absorption across the sample.

Similar statements can be made about the galactic size, shown in plot (b) of Fig. \ref{fig:vaucsource}. There is a mild trend of larger absorption with larger galactic size, which resembles what has been written in section \ref{sec:param}. However, the trend disappears after removing the three cases with the strongest absorption.

Reflecting the plot shown in Fig. \ref{fig:tau} (c), the absorption seems to decrease with increasing half-light radius, which is shown in Fig. \ref{fig:vaucsource} (c). As in the other cases, the trend depends strongly on the three galaxies with the largest absorption values. If these cases are removed, the dependence on the half-light radius vanishes.

Even with the three strong absorbers in, there is no trend between absorption and redshift as displayed in Fig. \ref{fig:vaucsource} (d). This is an important finding, since it tells us that our results are not biased by source selection. One might expect a trend on galactic evolution, but this is something which we do not take into account, and is not expected to have a strong influence for such small redshifts. 

On the other hand, it shows that the absorption value depends on the galactic parameters themselves. And given that there is no trend, or at most a mild trend with host parameters, the interplay between these parameters on galactic properties is striking and points to some fundamental relations of galactic evolution.

While the de Vaucouleurs' profile represents well the surface brightness of the outer parts of elliptical galaxies, its descriptive power of the inner parts where most of the absorption takes place is limited. Unfortunately, for no galaxy in our sample we have found more accurate profiles, and no conclusion can be drawn concerning the influence of the profile on the degree of absorption.

Tab. \ref{tab:deVauc} reveals that three sources, 1ES~0229$+$200, 1ES~0806$+$524, and 1ES~2344$+$514 exhibit starlight absorption in excess of $10\%$. Unfortunately, the uncertainties of the attenuation by the EBL are larger than these $20\%$. This means that the starlight absorption cannot be discerned from the EBL uncertainties in these sources until the EBL is known with higher accuracy. This is also true for AP Librae, where the starlight absorption could be useful to discriminate between different possible locations of the TeV emission region.

The highest degree of absorption by host galactic starlight in our sample is $\tau_{\gamma\gamma} = 0.201$. Taking into account the uncertainty in the EBL models, the absorption by the EBL starts to dominate the absorption by the host galaxy in sources with a redshift $z_r > 0.015$. Calculations by \cite{sea16} indicate that the EBL-absorption of $2\,$TeV photons from a source at $z_r=0.015$ is at a level of $\tau_{EBL}=0.15\pm0.06$, where the uncertainty is derived from the observational uncertainty of EBL measurements. The absorption by starlight dominates in sources with a smaller redshift. Most of the TeV emitting radio galaxies and all of the starburst galaxies emitting at these energies are found in this range of distances.

Employing a differential comparison of sources with comparable redshift ($\Delta z_r \lesssim 0.01 $) the starlight absorption can be revealed at larger redshifts, as well. Due to the small difference in redshift the uncertainties in the EBL model cancel for the compared sources. The TeVCat lists at least 8 pairs fulfilling the $\Delta z_r$ criterion. A potential pair of candidates are the blazars 1ES 0347-121 and 1ES 1218$+$304 at a redshift of z=0.182 and z=0.188, respectively. These blazars are intrinsically similar and exhibit a hard intrinsic spectrum. While the difference of the absorption values is modest ($\Delta \tau_{\gamma\gamma}\sim 0.02$), the high sensitivity of the Cherenkov Telescope Array \citep[e.g.,][]{aea08} should be able to detect this difference.

Apart from the potential detection of the starlight absorption in blazar spectra, our estimates of the degree of absorption in a large number of blazar hosts are important in order to determine the intrinsic properties of blazars. The modeling of observed spectra must take into account all absorption processes, and the absorption by starlight of up to $20\%$ should not be neglected.

%
%
\section{Summary \& Conclusions}
We have derived the general expression to calculate the attenuation of $\gamma$-ray photons by the optical starlight of elliptical galaxies by taking into account an arbitrary location and observation angle of the $\gamma$-ray emission region. The expression given in Eq. (\ref{eq:taugammagammaGEN}) is applicable also for other absorbing photon fields, and is therefore a general result. 

The precise degree of absorption depends on the distribution of the starlight photons within the galaxy. We used the de Vaucouleurs' profile, since its parameters are known for most galaxies. Deriving an average galaxy of a sample of blazar hosts, we conducted a parameter study on how different parameters of the $\gamma$-ray emission region and the host galaxy influence the absorption. Obviously, the absorption is strongest for TeV photons emitted in the central part of the galaxy, but even at a significant distance from the central region ($z\gtrsim 0.01 r_h$) the absorption is still $\sim 50\%$ of the central value. For an emission region beyond the half-light radius, the absorption is insignificant. 

We took into account the observation angle, because TeV emission has been observed in some AGN that are not blazars, where the observation angle could be much larger than the jet opening angle. The observation angle influences the absorption, since for $\muobs=1$ the TeV photons take the shortest route out of the galaxy causing less absorption than at any other angle. In fact, the absorption increases monotonically with increasing observation angle and is maximal for $\muobs=-1$. 

While keeping the total luminosity of the template galaxy constant, we varied the half-light radius and studied its influence on the degree of absorption. The absorption increases for smaller half-light radii, since more light is concentrated in the inner regions of the galaxy, which increases the density of the starlight photons. 

The influence of the galactic size describes the cumulative absorption TeV photons exhibit while traveling outward. The degree of absorption increases until the half-light radius, beyond which it reaches an asymptotic value. This is the same statement as above, namely that the additional absorption beyond the half-light radius is insignificant.

The benchmark galaxy gives an absorption of $\tau_{\gamma\gamma}\sim 0.06$. We analyzed the effect on theoretical spectra of blazars. We derived SSC spectra for different sets of parameters and applied the absorption. The effect is best detectable for blazars that exhibit a flat SED in the TeV range, such as HBL sources. The softer the spectrum in the TeV range, the more difficult is the detection of the absorption effect. Given that the maximum absorption is exhibited for photons of $\sim 2\,$TeV, the absorption is not detectable in sources cutting off at $\lesssim 1\,$TeV. Additionally, the spectral softening due to absorption by the EBL makes it more difficult to detect the influence of the host galaxy. 

We derived the maximum degree of absorption for a sample of host galaxies of TeV detected blazars. In most galaxies the degree of absorption is on the order of $10\%$ or less. We do not find significant correlations between absorption and characteristic parameters of the hosts, such as the emissivity, the size of the galaxy or the half-light radius of the brightness profile. This implies that these parameters are not independent from each other, pointing to fundamental relations for galaxy evolution. We also find no correlation between the degree of absorption and the redshift of the source, which excludes a strong selection bias in our sample.

An uncertainty in our results is the starlight profile. For most galaxies only the parameters of the de Vaucouleurs' profile are known. However, its predictive power of the central region of galaxies is limited, and more sophisticated profiles would be useful. The lack of these prevents a study on the influence of the profile on the absorption, and the explanatory power of averaged profiles that do not represent well individual galaxies, is at least questionable.

The uncertainties of the EBL inhibit the detection of the starlight absorption for any source beyond a redshift of $z_r = 0.015$. Nevertheless, a differential comparison of sources with similar redshift and similar intrinsic properties, but different hosts, could be used to detect the starlight absorption. The impact of starlight absorption on the TeV spectrum of the AGN is up to $20\%$, and this study has shown that the absorption by starlight in nearby TeV sources should be taken into account to construct the de-absorbed spectrum, which is used to derive the physical properties of the TeV emission region. 

%
%
\section*{Acknowledgement}
The authors wish to thank Lukasz Stawarz for valuable discussions, and the referee, Floyd Stecker, for a helpful report.
MZ and SJW acknowledge support by the German Ministry for Education and Research (BMBF) through Verbundforschung Astroteilchenphysik grant 05A11VH2. XC acknowledges support by the Helmholtz Alliance for Astroparticle Physics HAP funded by the Initiative and Networking Fund of the Helmholtz Association.
This research has made use of the NASA/IPAC Extragalactic Database (NED) which is operated by the Jet Propulsion Laboratory, California Institute of Technology, under contract with the National Aeronautics and Space Administration. 
%
%
\appendix
\section{Deriving the starlight emissivity} \label{app:emis}
The total luminosity $L$ of a galaxy can be calculated by integrating over the surface brightness distribution $S(r)$, giving
\begin{align}
L = 2\pi \intl_0^R S(r) r\td{r} \label{eq:applum}.
\end{align}
Following S06, the radial dependence function $h(r)$ can be written as
\begin{align}
h(r) = \frac{r_0}{r} \frac{S(r)}{S_0} \label{eq:apph},
\end{align}
with $S_0 = S(r_0)$, and $r_0$ being a characteristic radius. The monochromatic emissivity $j$ follows from the monochromatic luminosity $L$ of the host galaxy:
\begin{align}
j &= \frac{L}{(4\pi)^2 \int_0^R h(r) r^2 \td{r}} \nonumber \\
&= \frac{\int_0^R S(r) r \td{r}}{8\pi \frac{r_0}{S_0} \int_0^R S(r) r \td{r}} \nonumber \\
&= \frac{S_0}{8\pi r_0} \label{eq:appemis1} .
\end{align}

The surface brightness $\mu$ is typically measured in units of mag/arcsec$^2$. The transformation to solar luminosities per square-parsec is given by
\begin{align}
 F = 10^{0.4(M_{\odot}+21.572-(\mu-A))} \label{eq:appsurf1},
\end{align}
where $A$ is the Galactic extinction. Expressing $F$ in units of erg/cm$^2$/s yields $S_0$ and thus the emissivity in the R-band
\begin{align}
 j_R = 2.32\times 10^{-5} r_0^{-1} 10^{0.4\times(25.992-(\mu_R-A_R))} \label{eq:appemis2} .
\end{align}

%
%

%
%
\end{document}